\documentclass[11pt]{article}

\usepackage[final]{acl}

\usepackage{times}
\usepackage{latexsym}

\usepackage[T1]{fontenc}

\usepackage[utf8]{inputenc}

\usepackage{microtype}

\usepackage{inconsolata}

\usepackage{graphicx}
\usepackage{subcaption}
\usepackage{longtable}

\title{Proactive Conversational Assistant for Procedural Manual Tasks based on Audio and IMU}

 \author{Rehana Mahfuz, Yinyi Guo, Erik Visser, Phanidhar Chinchili \\
         Qualcomm Technologies, Inc.\\
         \texttt{\{rmahfuz, yinyig, evisser, phanich\}@qti.qualcomm.com}}
\begin{document}

\maketitle
\begin{abstract}
Real-time conversational assistants for procedural manual tasks often depend on video input, which can be computationally expensive and compromise user privacy. For the first time, we propose a real-time conversational assistant that provides comprehensive guidance for procedural manual tasks using only lightweight privacy-preserving modalities such as audio and IMU inputs from a user's wearable device to understand the context. Using a furniture assembly task and a cooking task, we show how this assistant proactively communicates step-by-step instructions to a user performing a procedural task, and answers user questions. We illustrate the data generation method and the system design to achieve such an assistant. 
On observing that an off-the-shelf language model is a talkative assistant but is not always able to answer questions correctly, we demonstrate how finetuning the model improves its ability to limit unnecessary dialogues with a 50\% increase in the precision, while also improving its ability to answer questions correctly, measured by a 150\% increase in the recall of answers. We further describe how such an assistant is implemented on an edge device with no dependence on the cloud.
\end{abstract}

\section{Introduction}

Mental load caused by the complexity of procedural manual tasks often makes people feel incapable of performing the task, even if they have the required physical ability. In industrial jobs, this necessitates training for new employees, which interferes with the tasks of experienced employees. For a layperson, this necessitates choosing between professional help and a few tedious iterations of trial and error. Video-based conversational assistants are experiencing slow adoption due to the privacy-compromising and heavy compute demanding nature of video. On-device processing of video necessitates bulky hardware, while uploading it to the cloud for processing compromises user privacy. To address this, we contribute:

\begin{itemize} 
\setlength\itemsep{-0.5em}

   \item The design of an on-device real-time proactive conversational assistant that relies on only audio and IMU inputs from a wearable device, hence significantly reducing computational complexity compared to video-based assistants.
   \item A method to generate a multi-turn conversation dataset with timestamps where a proactive assistant guides a user based on recognized activities while correcting mistakes and answering questions. 
   \item A finetuning method for the language model acting as an assistant that shows a 150\% improvement in recall of answers, a 50\% improvement in precision, and a 55\% improvement in the LLM-judged rating over its pretrained counterpart, averaged across tasks and model sizes.
\end{itemize}

\begin{figure*}[]
  \includegraphics[width=0.97\textwidth]{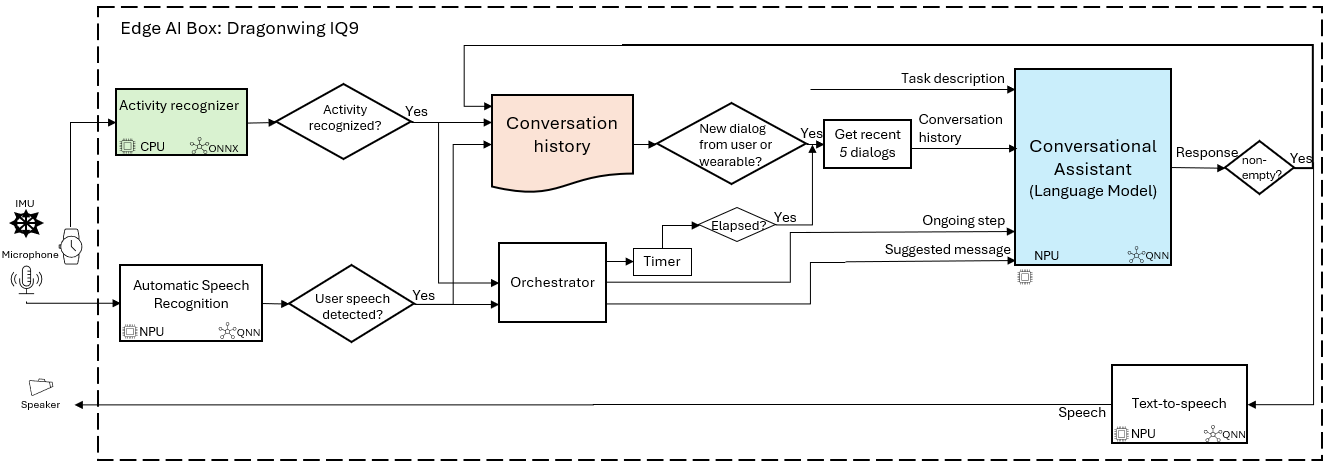}
  \caption{Design of our proactive situated conversational assistant. Occurrences of user comments or recognized activities trigger calls to the language model that provides responses as necessary.}
  \label{fig:overall_flowchart}
\end{figure*}

\section{Related Work}
The ubiquity of Multimodal Large Language Models (MLLMs) has enabled wearable assistants that maintain context awareness through an egocentric video stream \cite{install, svbench, videollm-online, basis, lion-fs, liu2024streamchat, wtag, morpheus, proactive}. While these are valuable while performing procedural activities, most offer only reactive assistance, i.e., the ability to answer questions. Only few \cite{morpheus,proactive} offer proactive assistance, i.e., narrating step-by-step instructions and intervening when needed. Moreover, they require the user to wear a Head Mounted Device (HMD) for egocentric video capture, which has limited adoption in daily lives because it can get uncomfortable to use for a long time.

On the other hand, smartwatches and smart rings have seen wider adoption due to their lightweight, comfortable and less obstructive nature. Various sensors from wrist-worn devices have potential in hand pose estimation, gesture recognition and activity recognition. \cite{pointing, qingxin2019unsupervised} used wrist IMU for activity recognition, while \cite{adaimi, samosa} used a combination of audio and IMU data. \cite{gestear, viband} achieved gesture recognition using wrist IMU, while \cite{viband, viobject} used wrist IMU for grasped object recognition. For hand pose estimation, \cite{echowrist} used wristband speakers, \cite{discoband} used depth sensors and \cite{hu2020fingertrak} used thermal cameras. Wrist-mounted cameras have shown further potential in developing context awareness \cite{wristsense, ohnishi2016recognizing, wacu}. 
While camera input is rich in information, it is largely redundant\cite{redundant}, apart from being large in volume.
\cite{prism-observer} developed a situated assistant relying on only IMU and audio inputs that intervenes during a procedural task to provide reminders for predefined activities and to correct mistakes, while \cite{prism-qa} showed how such an assistant can also answer questions if interfaced with an LLM on a server. However, these do not provide comprehensive guidance for a task, and need to be connected to a server to run the language model that acts as an assistant.
Ours is the first proactive conversational assistant that provides comprehensive step-by-step guidance using only lightweight modalities, running fully on the edge.

\vspace{-4mm}
\section{Method}
We illustrate in Figure \ref{fig:overall_flowchart} how a proactive conversational assistant can guide a user in performing a procedural task. Audio and IMU signals captured from a smartwatch worn on the dominant wrist are used to recognize activities performed by the user. The captured audio is transcribed into text to get the user's verbal comments and questions. Recognized user activities, user comments and questions along with the assistant's dialogues are logged in a conversation history document.
Every time an activity is recognized or the user says something or a timer elapses, an orchestrator prompts the language model to get any necessary messages, conveyed to the user as speech using a Speech-to-text model.

\subsection{Tasks for providing assistance}
\label{subsec: tasks}

\begin{figure*}[]
\begin{subfigure}[b]{0.65\textwidth}
  \includegraphics[width=0.65\textwidth]{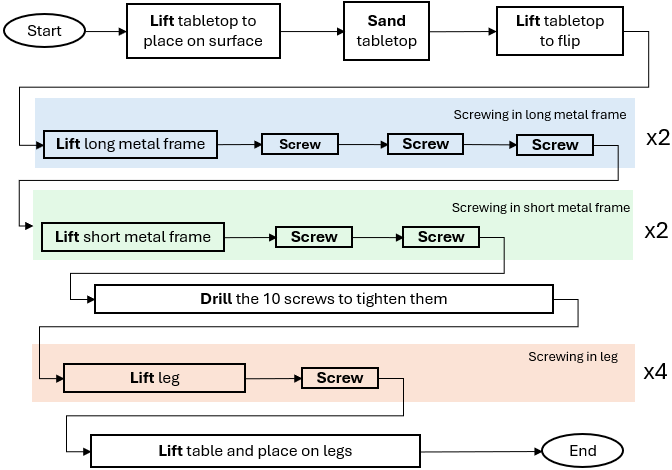}
  \caption{Furniture assembly task.}
  \end{subfigure}
  \hspace{-2cm}
  \begin{subfigure}[b]{0.65\textwidth}
  \includegraphics[width=0.65\textwidth]{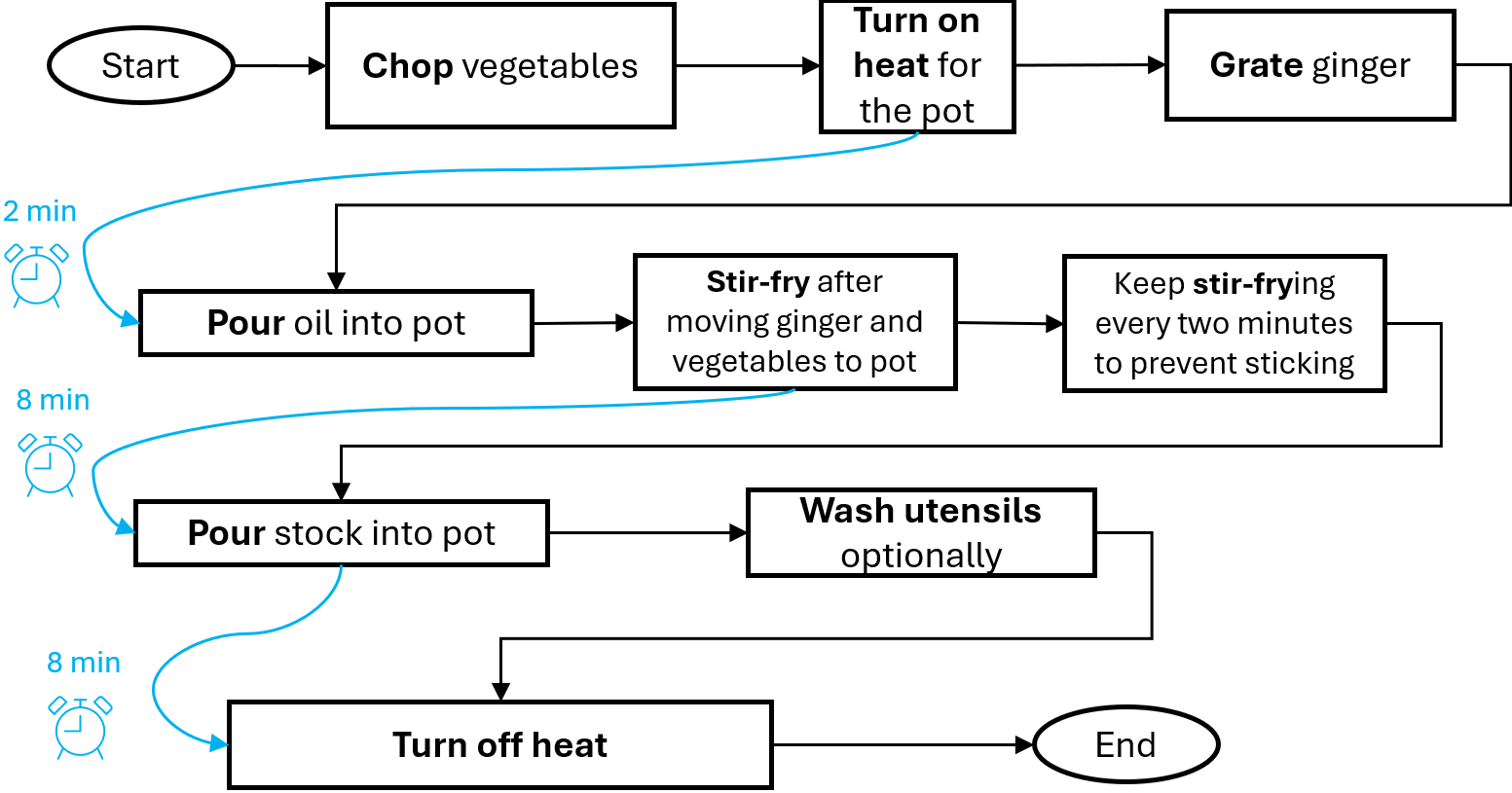}
  \caption{Cooking task.}
  \end{subfigure}

  \caption{Sequential steps for the procedural tasks.}
  \label{fig:tasks}
\end{figure*}

In the furniture assembly domain, we consider a table assembly task loosely based on \cite{ikea_sandsberg_assembly} and in the cooking domain, we consider the task of making soup, loosely based on \cite{plant-based-jess-soup-recipe}.
The table assembly task involves sanding a tabletop placed on a surface followed by flipping it, screwing in and drilling four metal frames and then screwing in four legs. Frames are screwed in vertically, while the legs are screwed in horizontally. Possible mistakes considered are not screwing in a frame immediately after placing it, drilling before placing all screws for frames, and screwing in legs before drilling all frames.
The soup making task starts with chopping vegetables, heating a pot and grating ginger. Oil is poured two minutes after turning on heat, followed by adding the chopped and grated vegetables and stirring them at least every two minutes. Vegetable stock is poured eight minutes after adding vegetables to the pot, after which the user can optionally wash utensils. The heat is turned off eight minutes after pouring the stock. Possible mistakes considered are forgetting to turn on heat before grating the ginger, forgetting to stir every two minutes, and starting to pour stock prematurely.

The two tasks are complimentary in nature. While table assembly involves keeping count of repeated activities like lifting, screwing and drilling and tracking their order, making soup is sensitive to performing activities at the correct times.
The full sequence of steps for both tasks are illustrated in Figure \ref{fig:tasks}.

\subsection{Activity Recognition}
Audio and IMU signals captured by a smartwatch worn on the user's dominant wrist are input into a neural network to recognize activities. For furniture assembly, the activities considered are sanding, drilling, screwing, unscrewing and lifting. For the first three activities, the data is sourced from the SAMoSA dataset \cite{samosa}. The data for the unscrew activity was collected by the authors, and has 61 and 10 samples for training and validation respectively. The data for the lifting activities was sourced from the BoxLift dataset \cite{boxlift}. 
For cooking, the activities considered are turning on heat, turning off heat, chopping, grating, pouring, stir-frying and washing utensils. The data for the first two activities correspond to screwing and unscrewing from the furniture assembly task. Chopping, grating and washing utensils are sourced from both the SAMoSA and the Adaimi \cite{adaimi} datasets, while pouring is taken only from SAMoSA and stir-frying is taken only from the Adaimi dataset. Of all available data, 85\% was used for training and 15\% was used for validation, unless specified otherwise. All datasets have audio and IMU data from a wrist-worn device, except the BoxLift dataset, which has only IMU data. While the labeled data does not distinguish between horizontal and vertical (un)screwing, we did so by observing that the z-axis peaks and troughs in the acceleration signal are more pronounced for vertical (un)screwing than for horizontal.

The neural network contains a dedicated encoder for each of the two modalities, followed by a linear layer for classification. It outputs frame-by-frame probabilities for each class, with each frame roughly corresponding to a second.
For the audio encoder, we used a CNN10 PANN \cite{cnn10} which has four convolution layers followed by a fully connected layer. For the IMU encoder, we used an Attend\&Discriminate model \cite{adaimi} which has four convolution layers, followed by self attention, two recurrent layers and finally temporal attention. All inputs are either zero-padded or truncated to achieve a 10 second input length. All audio and IMU inputs were resampled to 32 kHz and 50 Hz respectively. For the lifting activities, a tensor consisting of zeros was used for the audio input.
This neural network was trained for 200 epochs with a batch size of 8 and a learning rate of 0.001. The checkpoint with the highest F-score was selected.


\subsubsection{Conversation Data generation}
For each task, a rule-based conversation generator creates a conversation by randomly choosing a start time between 9 am and 5 pm and adding dialogues from the assistant, the wearable and the user, based on the predefined sequence of steps. Each dialogue is accompanied by a timestamp from a time tracker variable that is incremented each time a dialogue is added. For user or assistant dialogues, it is incremented by one second, and for dialogues from the wearable, it is incremented by a predefined duration of the activity, with some variance added randomly.
Occasionally, a user comment such as `Okay' following an instruction or `Okay, done' following an activity completion is added.
Based on a given skill level of a user between 0 and 1, mistakes mentioned in \ref{subsec: tasks} are added if the output of a random number generator exceeds the skill level. The locations of added mistakes is shown in Figure \ref{fig: mistake-insert}. Occasionally, the user challenges the mistake correction of the assistant, after which the assistant provides further explanation. Sometimes the user chooses to rectify their mistakes, or at least verbally acknowledge them. Occasionally, the user also takes a break by verbally indicating it. 
Sometimes the user inquires about the time remaining, in which case the assistant responds by subtracting the estimated time to finish from the current time.

All randomness is created by thresholding the output of a random number generator. Dialogues were paraphrased to introduce diversity.
We generated 1000 conversations for each task, split into sets of 900 and 100 for training and validation respectively. The distribution of skill levels is uniform.


\subsection{Conversational Assistant}
\label{subsec: conv_assist}

Conversational assistance is provided by a language model which is prompted each time an activity is recognized, the user has a comment or a question, or a timer for a time-sensitive step elapses. The prompt contains a description of the procedural task the user is performing, some helpful information from an orchestrator, and the five most recent dialogues from the conversation history. Older dialogues are omitted to limit the prompt size so that the delay during inference is minimal. 
If there is nothing informative to be conveyed, the language model is expected to produce an empty response.

A task-specific orchestrator maintains counters for each recognized activity and timestamps for some activities to determine which step the user is on, and provides this step information along with an optional suggested message to the language model in the prompt. This suggested message could be a key instruction, a mistake correction message, an answer or a miscellaneous message. The orchestrator also classifies the message into one of these types. 
For example, in the furniture assembly task, when the counters indicate the completion of six screws, the orchestrator prompts the language model with step information indicating that two long frames have been screwed in, and a suggested message indicating that the next step is to place a short frame followed by screwing it in with two screws. After this key instruction is conveyed, if the user disobeys and decides to drill instead, the orchestrator prompts the language model suggesting that the user started drilling prematurely and a mistake correction message suggesting that the user first finish screwing in all ten screws before drilling. If the user asks why, the orchestrator does not have an answer, and we rely on the language model to formulate an answer using information from the notes in its prompt or its knowledge from finetuning. This shows how we leverage the orchestrator for step tracking, and the language model for messages that require more information than what the orchestrator has.

In the cooking task, the orchestrator starts a timer if needed, and prompts the language model when the timer elapses. For example, when stir-frying begins, the orchestrator records this timestamp and starts an eight minute timer. If the user inquires about the remaining time, the orchestrator subtracts the time at which the question is asked from the estimated time to finish and conveys that. When this timer elapses, the language model is prompted to let the user know that it is time to pour stock. 



\subsection{Preparing the Language Model}

From the Qwen2.5 \cite{qwen} set of models, we experimented with the following three language models as assistants: Qwen2.5-0.5B-Instruct, Qwen2.5-1.5B-Instruct and Qwen2.5-3B-Instruct. We evaluated the ability of the pretrained models to act as assistants, and then finetuned them.
\subsubsection{Finetuning}
To finetune the language model, we performed Low Rank Adaptation (LoRA) of the model for two epochs with rank 8, alpha 16, batch size 8, dropout 0.075, and learning rate 3e-5. The checkpoint with the highest F-score was selected. 

The model is finetuned to match the assistant's ground truth dialogues, including empty responses. Sometimes when the user makes trivially informative whimsical statements such as `Okay' and `What next?', the ground truth conversation contains a non-empty assistant dialogue only after the whimsical statement. Because the user is not guaranteed to make such a statement, we make an adjustment by finetuning the model to match the next assistant dialogue even before the user makes the whimsical statement. We found this to be an important adjustment to preserve the model's proactive nature, and refer to this adjustment as User Whim Agnostic (UWA) finetuning.

\subsection{Metrics}
\label{sec: metrics}
To understand the assistant's strengths and weaknesses, the orchestrator categorizes assistant-generated dialogues into four types: key instruction, mistake correction, answer and miscellaneous.
Key instructions are the instructions needed by the user to perform the task correctly, even when they do not make any mistakes.
Mistake correction dialogues are the corrective statements the assistant generates if the user is making mistakes.
Answers are responses immediately following a user comment or question, if they do not fall in the previous two categories.
All other assistant-generated responses are categorized as miscellaneous. Some examples of miscellaneous statements are encouragements and periodic reminders.

To measure how correct an assistant-generated response is with respect to a ground truth dialogue, we measured the SentenceBERT score and the BERTScore between the two, as well as the entailment of the assistant-generated response to the ground truth dialogue. To measure these, we used all-MiniLM-L6-v2 model from the SentenceTransformers \cite{sbert} library, the roberta-large \cite{roberta-large} model from the bert-score \cite{bert-score} library, and the nli-debertav3-base \cite{debertav3} model from the SentenceTransformers library respectively.

In addition, we adopted the LLM-as-a-judge \cite{llm-as-a-judge} framework using GPT-5 \cite{gpt-5}. Further details are in Appendix \ref{appendix: metric_details}.
To measure the recall and precision for each type of assistant response, similar to \cite {semantic_entropy}, we consider a response as being correct with respect to the ground truth dialogue if either the SentenceBERT score exceeds 0.3 or the entailment score exceeds 0. The BERTScore was not used since it does not vary much.

\subsection{On-device Implementation Details}
All processing including the activity recognition, orchestration, language model inference and conversion between text and speech occur on an Edge AI box with a Dragonwing IQ9 \cite{iq9} processor.
A custom Android application on a watch with a Snapdragon W5 Gen 1 processor streams audio and IMU data at 32kHz and 50 Hz respectively to the edge device through a Redis-based microservice \cite{redis}. Here, these time series signals are input into the activity recognition neural network exported to ONNX and deployed using an onnxruntime session.

The user's speech streamed from the watch is transcribed into text using Whisper‑medium \cite{whisper}. The assistant's response is converted to speech using the MeloTTS-English \cite{melotts} model. The resulting audio packets are streamed back to the smartwatch using a GStreamer-based service, enabling low‑latency spoken feedback.
All models on this edge device except activity recognition are run as dedicated microservices in Docker containers on the Neural Processing Unit by converting to the QNN format using the Qualcomm® AI Engine Direct SDK \cite{qairt-sdk}. 

\section{Results}


\subsection{Conversation Data Quality Evaluation}

\begin{figure}[]
\includegraphics[width=0.41\textwidth]{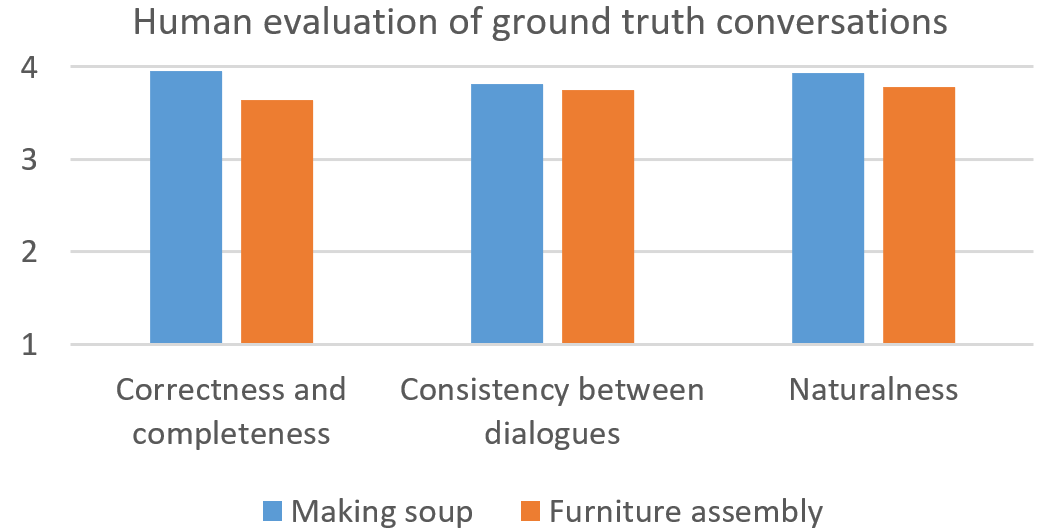}
  \caption{Human evaluation of the dataset quality.}
  \label{fig: dset_human_eval}
\end{figure}


Seven human subjects evaluated four ground truth conversations from each task on three metrics on a scale of 1-4, as shown in Figure \ref{fig: dset_human_eval}. While the result is reasonable, evaluators penalized instances where the user disregarded the assistant's instructions even when they were clear, and instances where the assistant did not try a different paraphrasing of the mistake correction message when mistakes were repeated. The user's disregard of the assistant's instructions was deliberately introduced to simulate user sloppiness. Paraphrasing mistake correction messages when ineffective will be considered in future work.

\subsection{Activity Recognition}

\begin{figure}[]
\includegraphics[width=0.47\textwidth]{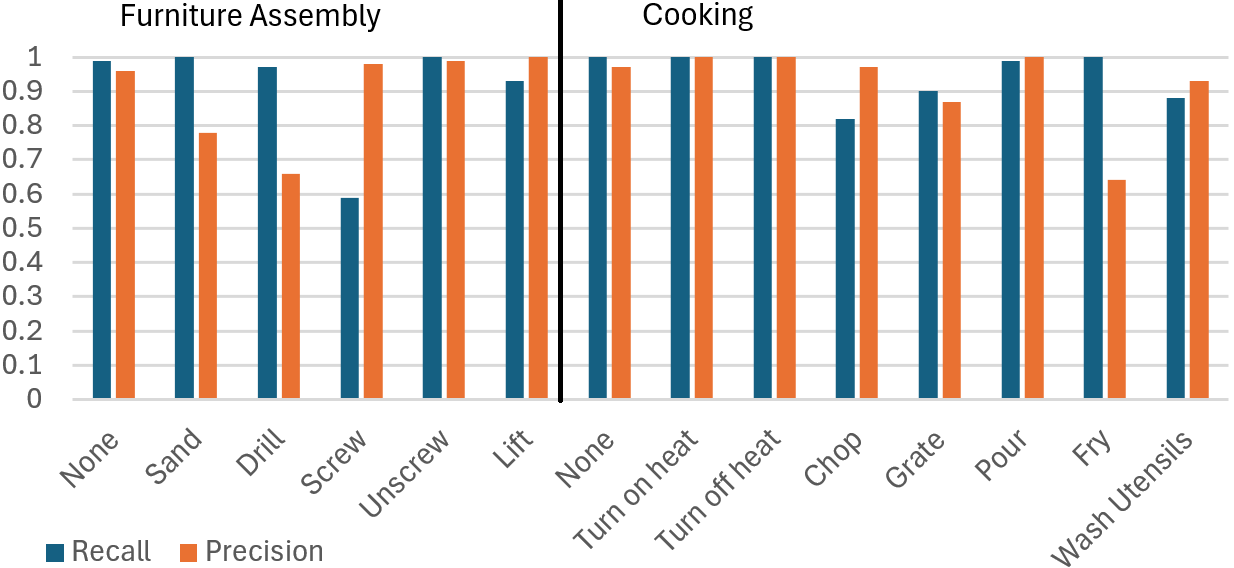}
  \caption{Activity recognition classification result.}
  \label{fig: act-rec}
\end{figure}

Figure \ref{fig: act-rec} shows the validation recall and precision of the trained activity recognizer.
Screwing is sometimes mistaken as sanding or drilling, explaining its low recall. Stir-frying sometimes gets detected as chopping, explaining its low recall and the low precision for frying. These mistakes negatively impact the ability of the assistant to follow the user's actions, which impacts its ability to guide the user. To address these mistakes, determining class-specific probability thresholds may be helpful instead of simply selecting the most confident class label.


\subsection{Conversational Assistant}
\begin{figure}[]
\includegraphics[width=0.48\textwidth]{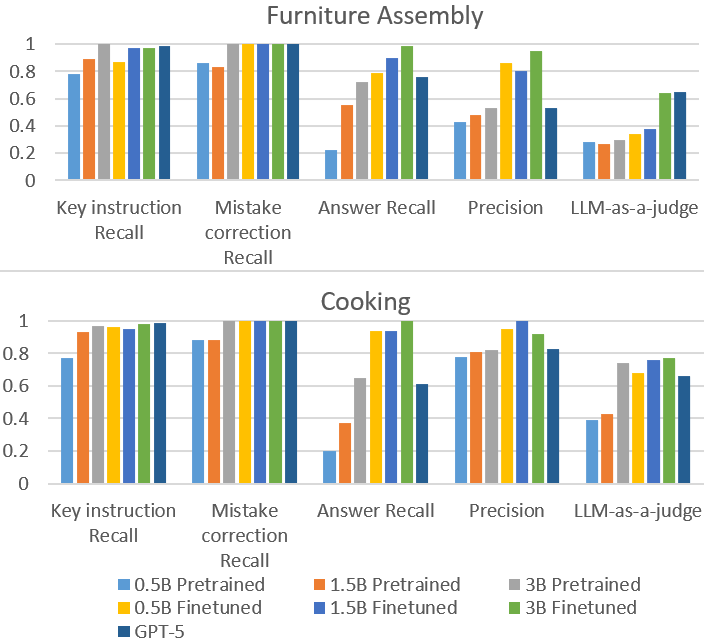}
  \caption{Performances of different language models as assistants for both tasks.}
  \label{fig: main_result}
\end{figure}

Figure \ref{fig: main_result} shows the performances of models of all three sizes for both tasks before and after finetuning. 
Since the key instructions and possible mistakes are clearly mentioned in the prompt, the pretrained models are able to deliver the key instructions and mistake correction messages satisfactorily. This is especially true as the model size gets bigger. Answering user questions is more challenging, and the pretrained models struggle here, even though most answers are present in the system prompt. Additionally, the pretrained models sometimes repeat messages unnecessarily or hallucinate by generating irrelevant messages such as `Let's proceed with the steps as instructed.' and `Let me help you with that. Here we go.', as measured by the precision.
The precision for the furniture assembly task is lower than that for the cooking task because the former has more steps and hence more instances where no message is needed from the assistant.

Finetuning the models improved the assistant's ability to answer questions, while maintaining and sometimes slightly improving the ability to deliver key instructions and mistake correction messages. Further, finetuning also conditioned the model to limit less relevant messages. The imperfect precision after finetuning is caused by messages which are relevant but not strictly necessary. Some examples are warnings against mistakes even before they are made, reinforcements after mistake corrections such as `This way, everything tastes better.' and suggestions to make the user feel free such as `You can now go do other things, knowing that by 2:08 PM, the soup will be done.'.

To test the limits of how well a pretrained model can perform, we tried using GPT-5 as an assistant. Quantitatively, its responses pass the thresholds of our correctness measures for most messages except some answers. This is because answers can be open ended. Qualitatively, its messages are long and verbose, which may be appealing to users looking for detail, but not to users looking for objective guidance. It also makes a lot of relevant but unnecessary comments such as `I'll wait while you finish that'. Another example is that it keeps encouraging the user after each drilling activity with a message like `Looks good—keep tightening the remaining screws with the drill until they’re all snug. Don’t over-tighten.'. In contrast, our finetuned models offer succinct objective guidance while being small enough to be deployed on the edge.




\section{Conclusion}
We illustrate a proactive conversational assistant that guides a user step-by-step through procedural tasks, while leveraging audio and IMU data from a wearable device to maintain context awareness. Through both similarity and entailment metrics as well as using the LLM-as-a-judge framework, it is established that finetuning a language model improves its ability to provide objective guidance while limiting less relevant messages. Future work will involve automating enrollment of new tasks to this framework, as well as exploring tool-calling as an addition or as an alternative to finetuning.

\section{Limitations}
\begin{itemize}
\item Our activity recognizer assumes that the user's dominant wrist is their right wrist. This holds true for only 90\% of the world population \cite{right-hand}.
\item When deployed in a setting outside of the dedicated task, the activity recognizer may trigger false alarms. For example, the turn of a doorknob, while unlikely while making soup, may be detected as turning the knob of a cooking stove to turn on heat. Further training of the activity recognizer with negative examples of common false alarms would address this.
\item While this work tries to push the limits of how effective an assistant can be using only lightweight privacy-preserving sensors, its context awareness capability is limited compared to  that of assistants using richer modalities such as video. For example, if the user chopped radishes instead of carrots, it would be easier to determine that using video than using audio and IMU. Distinguishing between vegetables being chopped using audio and IMU is still possible because of the differences in the firmness of vegetables.
\item This system does not have safety guardrails to ensure the user is not hurting themselves during the task. For example, if the tabletop falls on the user's feet and hurts them, or if violently spluttering oil while frying burns the user's skin, the system cannot prevent it beyond warning the user to be careful.
\item The quality of the reported metrics is subject to the quality of the models used to calculate them.
The recall and precision depend on the SentenceBERT and entailment models used to judge the correctness of a generated dialogue with respect to a reference dialogue. The LLM-as-a-judge metric depends on GPT-5 to judge the quality of the assistant's guidance.
\item For a particular task, a language model finetuned for a different task doesn't perform better than a pretrained model. To obtain a model that performs well across tasks, it may need to be finetuned with all tasks.

\end{itemize}




\bibliography{custom}

\appendix

\section{Details of Metrics}
\label{appendix: metric_details}

Table \ref{table: llm-as-a-judge-prompt} shows the system prompt provided to GPT-5 to judge the quality of the assistant's guidance with respect to an exemplary ground truth conversation. The rating was divided by 10 before reporting to ensure it lies between 0 and 1 like the recall and precision do.

\begin{table*}[]
    \centering
    \begin{tabular}{|p{14.5cm}|}
    \hline
You are an impartial judge who is evaluating the quality of responses provided by an AI assistant to a user who is <assembling a table/making soup>. \\
Here are the instructions to <assemble a table/make soup>: \\

<instructions including materials, steps and notes> \\

First you will be given a reference conversation where the assistant's responses are exemplary. Then you will be given the conversation that occurred with the actual assistant. Your task is to evaluate the quality of the actual assistant's responses based on correctness, helpfulness and restraint. Restraint refers to the assistant's ability to avoid saying something unless necessary.\\
Please provide one overall rating for all responses on a scale of 1 to 10 by strictly following this format: "    [[rating]]", for example: "Rating: [[5]]
\\ \hline
    \end{tabular}
   \caption{System prompt provided to the language model for judging the quality of an assistant's guidance.}
    \label{table: llm-as-a-judge-prompt}
\end{table*}


\section{Prompts for Language Models}
\label{appendixA}
Tables \ref{table: sandsberg_sys_prompt} and \ref{table: cook_sys_prompt} show the system prompts provided to the language model acting as an assistant for the furniture assembly and cooking tasks respectively.

\section{Example Conversations}
\begin{figure*}[]
  \includegraphics[width=14.5cm]{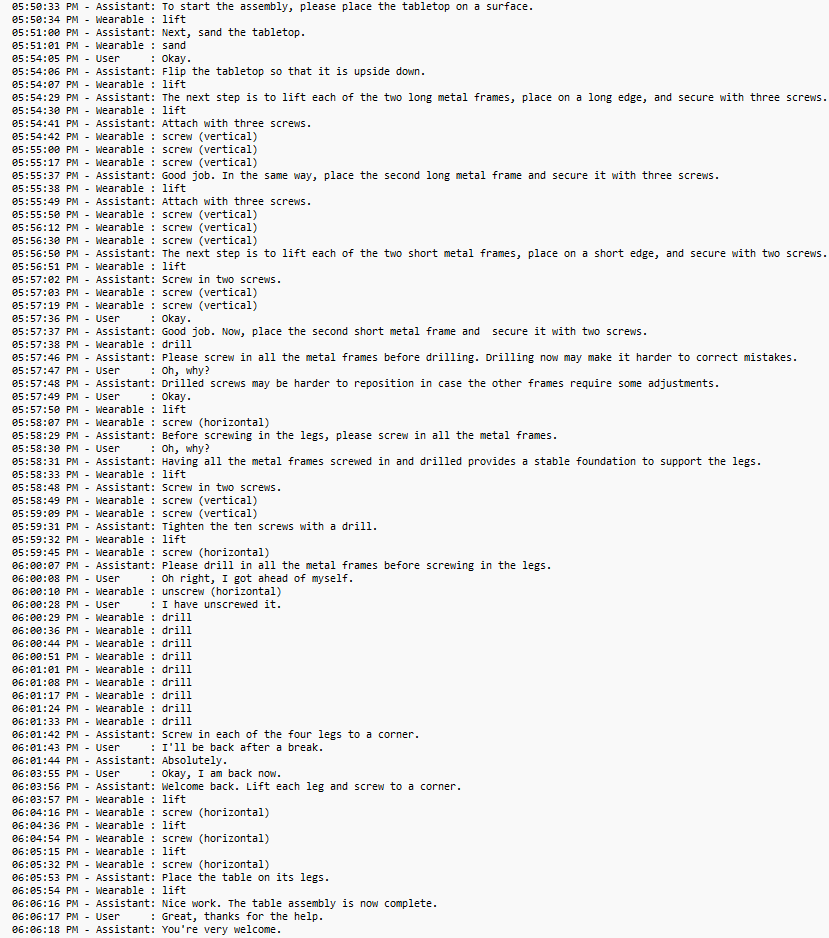}
  \caption{Example ground truth conversation for the furniture assembly task.}
  \label{fig: sandsberg_ex-convo}
\end{figure*}

\begin{figure*}[]
  \includegraphics[width=14.5cm]{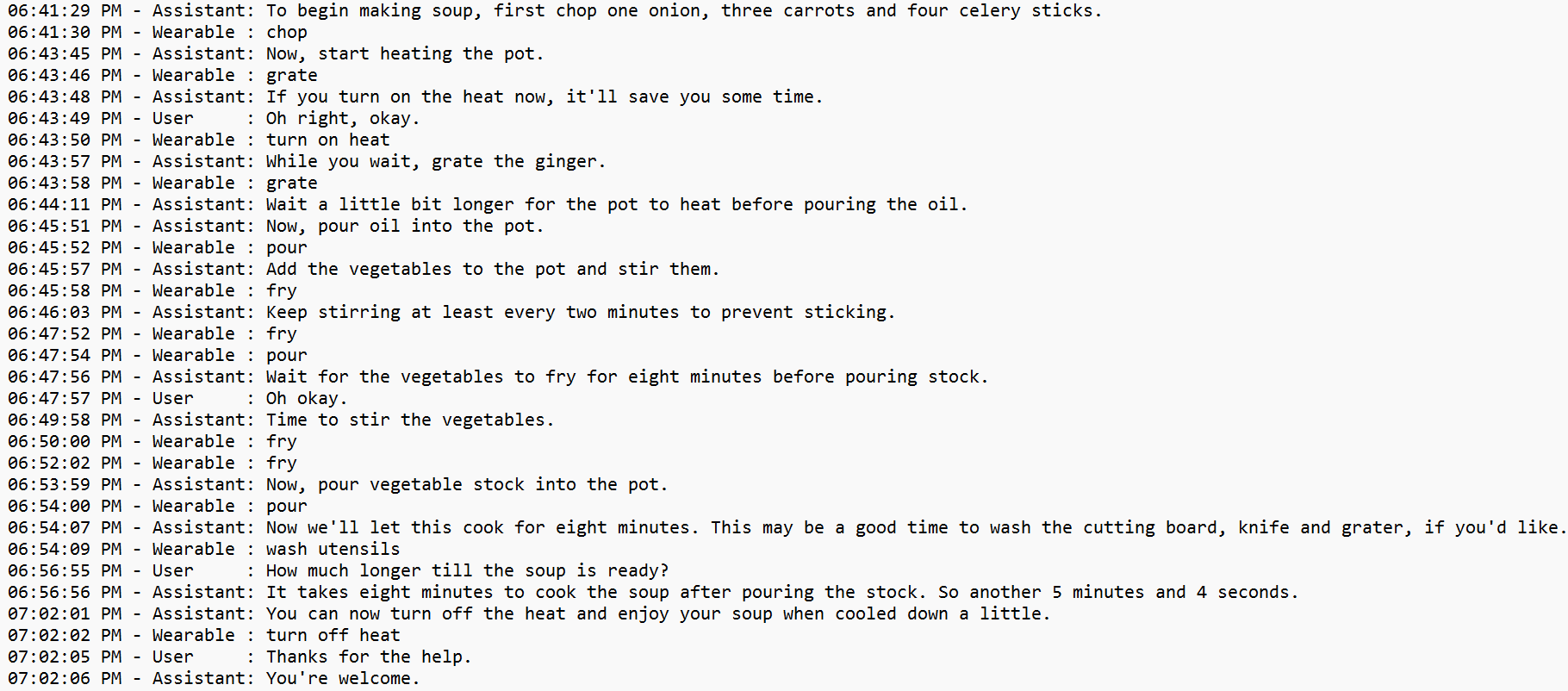}
  \caption{Example ground truth conversation for the cooking task.}
  \label{fig: cook_ex-convo}
\end{figure*}

\begin{figure*}[]
  \includegraphics[width=14.5cm]{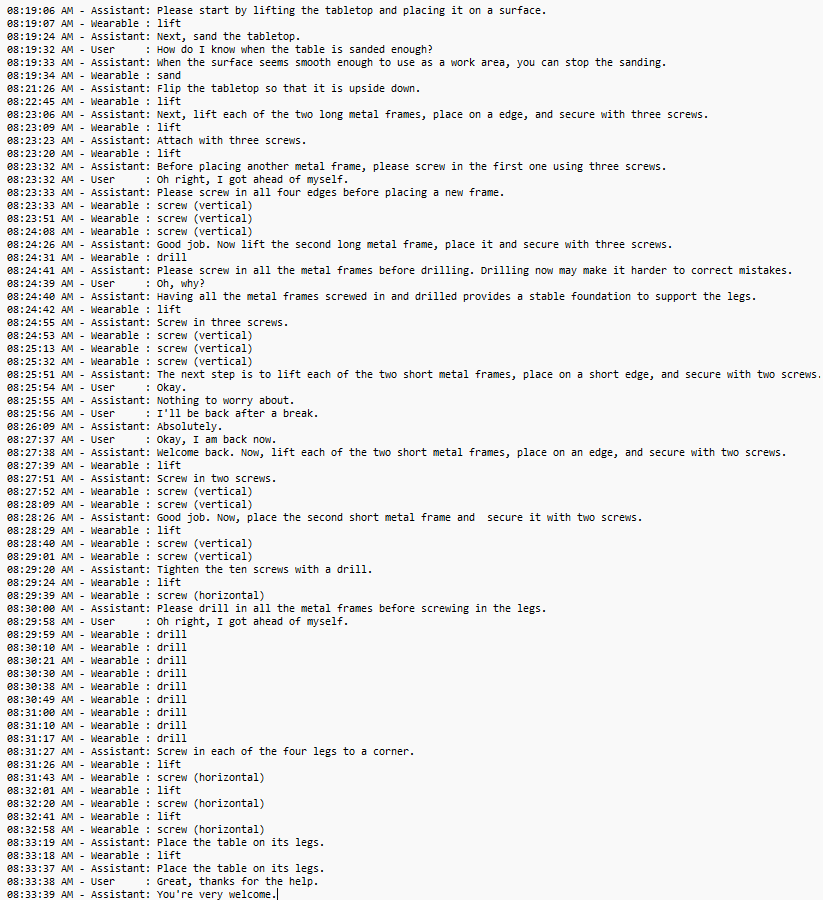}
  \caption{Example conversation where the finetuned 3B model acts as an assistant for the furniture assembly task.}
  \label{fig: sandsberg_3B_ex-convo}
\end{figure*}

\begin{figure*}[]
  \includegraphics[width=14.5cm]{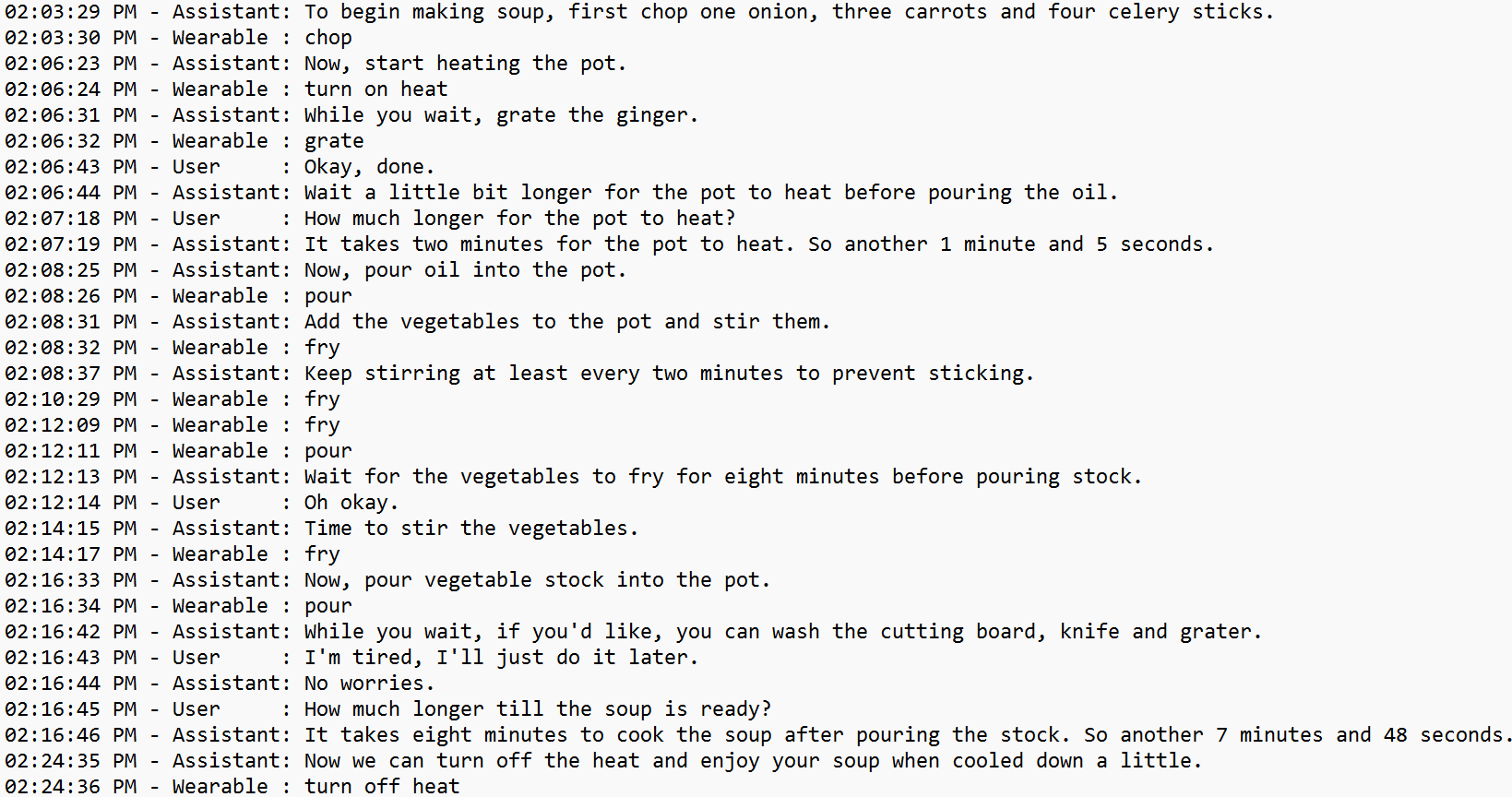}
  \caption{Example conversation where the finetuned 3B model acts as an assistant for the cooking task.}
  \label{fig: cook_3B_ex-convo}
\end{figure*}

Figures \ref{fig: sandsberg_ex-convo} and \ref{fig: cook_ex-convo} show example ground truth conversations for the furniture assembly and cooking tasks respectively.
Figures \ref{fig: sandsberg_3B_ex-convo} and \ref{fig: cook_3B_ex-convo} show conversations where a finetuned 3B model acts as an assistant.

\begin{table*}[]
    \centering
    \begin{tabular}{|p{14.5cm}|}
    \hline
        Please assist the user in assembling a table step-by-step, given the assembly manual. At each conversation turn, you will be given information about which step they are on. \\

Materials: tabletop, 4 metal frames, 4 legs

Steps:

Step 1: Prepare the tabletop \\
\hspace{1cm}Step 1.1: Lift the tabletop and place it on a surface.\\
\hspace{1cm}Step 1.2: Sand the tabletop.\\
\hspace{1cm}Step 1.3: Lift the tabletop again and place it upside down.\\

Step 2: Attach the metal frames\\
\hspace{1cm}Step 2.1: Lift each of the two long metal frames, place on an edge, and secure with three screws.\\
\hspace{1cm}Step 2.2: Lift each of the two short metal frames, place on an edge, and secure with two screws.\\
\hspace{1cm}Step 2.3: Tighten all 10 screws using a drill.\\

Step 3: Attach the legs\\
\hspace{1cm}Step 3.1: Lift each of the four legs leg and screw it to a corner of the tabletop.\\
\hspace{1cm}Step 3.2: Lift the table and place it on its legs.\\

Notes: \\
1) Once a metal frame is placed, to prevent it from getting moved by mistake, it must be screwed in completely before placing another metal frame.\\

2) Steps 2.1 and 2.2 (screwing metal frames) involve vertical screwing. Step 3.1 (screwing legs) involves horizontal screwing. A screwdriver is recommended for initial screwing over a drill, as a screwdriver gives better control.\\

3) Before starting to drill, all four frames must be screwed in fully using ten vertical screws. Drilling before securing all vertical screws makes it harder to adjust frames that are already drilled in, as other frames are being secured.\\

4) The legs should only be screwed in after screwing and drilling in the metal frames, since these frames provide a stable base for the legs.\\

5) If the user wants to take a break, allow them to do so without talking about the next step till they return.\\

6) Keep your responses short. If anything unexpected occurs, use your best judgement.\\

7) Please only generate the assistant's dialog, and not the dialogs of the wearable or the user.\\

8) Do not refer to the numbers of the steps or notes, as the user has no knowledge of these.\\

\\ \hline
    \end{tabular}
   \caption{System prompt provided to the language model for the furniture assembly task.}
    \label{table: sandsberg_sys_prompt}
\end{table*}

\begin{table*}[]
    \centering
    \begin{tabular}{|p{14.5cm}|}
    \hline
Please assist the user in making soup step-by-step, given a list of sequential steps. At each conversation turn, you will be given information about which step they are on. \\

Materials for making soup: 1 onion, 3 carrots, 4 celery sticks, ginger, vegetable stock, oil, pot \\

Steps:\\
1)  Chop one onion, three carrots and four celery sticks.\\
2)  Turn on heat for the pot.\\
3)  Grate ginger.\\
4)  Pour oil in the pot after two minutes of heating the pot.\\
5)  Add vegetables to the pot and stir (detected as frying).\\
6)  Keep stirring at least every two minutes to prevent sticking.\\
7)  After eight minutes, pour vegetable stock into the pot.\\
8)  Optional: wash the cutting board, knife, grater etc. while waiting.\\
9)  Turn off the heat after eight minutes of pouring the stock.\\

Notes:\\
1)  If user doesn’t turn on heat before starting to grate ginger, they’ll have to wait longer for the pot to heat later.\\

2) If the user performs steps out of order, please correct them.\\

3) Keep your responses short. If anything unexpected occurs, use your best judgement.\\

4) Please only generate the assistant's dialog, and not the dialogs of the wearable or the user.\\

5) Do not refer to the numbers of the steps or notes, as the user has no knowledge of these.\\

\\ \hline
    \end{tabular}
   \caption{System prompt provided to the language model for the cooking task.}
    \label{table: cook_sys_prompt}
\end{table*}

\section{Mistake insertion during Data Generation}
\begin{figure*}[]
\begin{subfigure}[b]{0.99\textwidth}
  \includegraphics[width=0.99\textwidth]{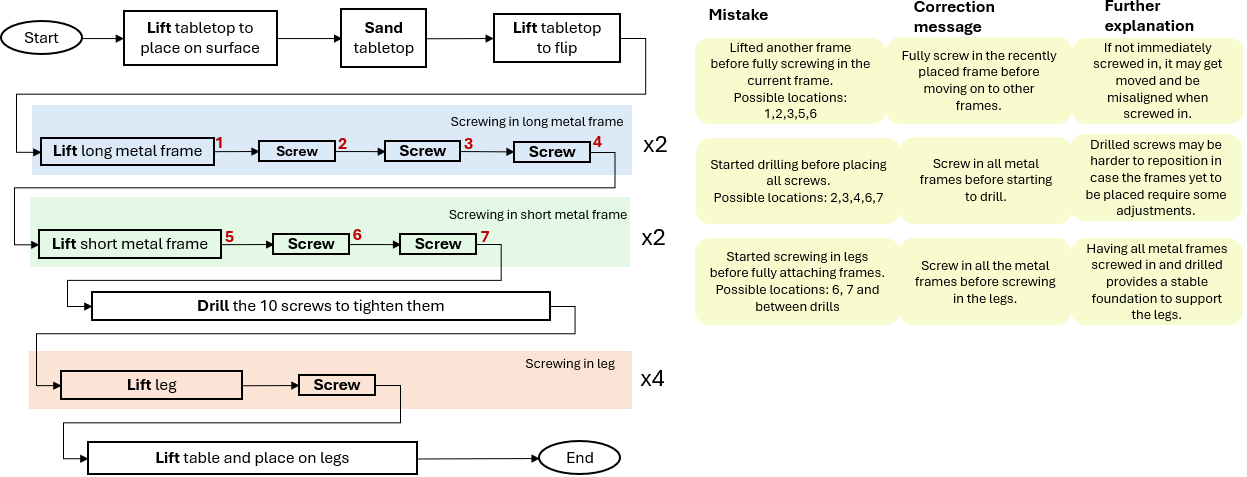}
  \caption{Furniture assembly task.}
  \end{subfigure}
  \begin{subfigure}[b]{0.99\textwidth}
  \includegraphics[width=0.99\textwidth]{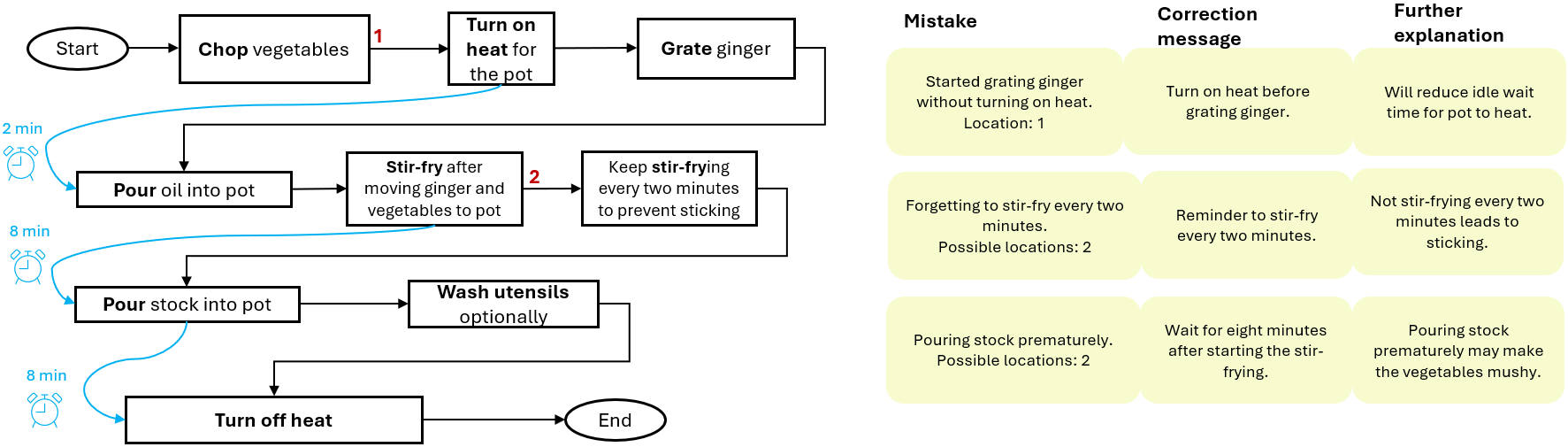}
  \caption{Cooking task.}
  \end{subfigure}

  \caption{Mistake insertion during conversation generation.}
  \label{fig: mistake-insert}
\end{figure*}

Figure \ref{fig: mistake-insert} shows where mistakes are inserted during conversation generation.

\end{document}